\documentclass[twocolumn,showpacs,preprintnumbers,amsmath,amssymb]{revtex4}

\usepackage{graphicx}
\usepackage{dcolumn}
\usepackage{bm}
\usepackage{epsfig}
\usepackage{amssymb}
\usepackage{amsthm}
\usepackage{verbatim}
\usepackage{delarray}
\usepackage{graphics}
\usepackage{epic}

\newcommand{\blc}{\left\{}
\newcommand{\brc}{\right\}}

\newcommand{\ket}[1]{{|#1\rangle}}
\newcommand{\bra}[1]{{\langle#1|}}

\newcommand{\bl}{\left(}
\newcommand{\br}{\right)}

\newcommand{\tr}{\mbox{Tr}}

\newcommand{\deft}{\ {\stackrel{\triangle}{=}} \ }

\newcommand{\hpi}{\widehat{\Pi}}

\newcommand{\hz}{\widehat{Z}}

\newcommand{\LL}{{\mathcal{L}}}

\newcommand{\B}{{\mathcal{B}}}

\newcommand{\HH}{{\mathcal{H}}}

\newcommand{\G}{{\mathcal{G}}}
\newcommand{\I}{{\mathcal{I}}}

\newcommand{\SSS}{{\mathcal{S}}}
\newcommand{\K}{{\mathcal{K}}}

\newcommand{\ie}{{\em i.e., }}
\newcommand{\eg}{{\em e.g., }}

\newcommand{\etal}{\emph{et al.\ }}

\newtheorem{theorem}{Theorem}

\begin{document}


\title{Optimal quantum detectors for unambiguous detection of mixed states}
\author{Yonina C. Eldar}
\affiliation{Department of Electrical Engineering,
Technion---Israel Institute of Technology, Haifa 32000, Israel}
 \email{yonina@ee.technion.ac.il}
\author{Mihailo Stojnic}
\affiliation{Department of Electrical Engineering, California
Institute of Technology, Pasadena, CA 91125, USA}
\email{{mihailo,hassibi}@systems.caltech.edu}
\author{Babak Hassibi}
\affiliation{Department of Electrical Engineering, California
Institute of Technology, Pasadena, CA 91125, USA}
\email{hassibi@systems.caltech.edu}
\date{\today}

\begin{abstract}

We consider the problem of designing an optimal quantum detector
that distinguishes unambiguously between a collection of mixed
quantum states. Using arguments of duality in vector space
optimization, we derive necessary and sufficient conditions for an
optimal measurement that maximizes the probability of correct
detection. We show that the previous optimal measurements that
were derived for certain special cases satisfy these optimality
conditions. We then consider state sets with strong symmetry
properties, and show that the optimal measurement operators for
distinguishing between these states share the same symmetries, and
can be computed very efficiently by solving a reduced size
semidefinite program.

\end{abstract}

\maketitle

\section{Introduction}

The problem of detecting information stored in the state of a
quantum system is a fundamental problem in quantum information
theory. Several approaches have emerged to distinguishing between
a collection of non-orthogonal quantum states. In one approach, a
measurement is designed to maximize the probability of correct
detection
\cite{H73,H76,EMV02,CBH89,OBH96,BKMO97,EF01,EMV02s,EF01,E03l}. A
more recent approach, referred to as unambiguous detection
\cite{I87,D88,P88,JS95,PT98,C98,CB98,E02,E03p}, is to design a
measurement that with a certain probability returns an
inconclusive result, but such that if the measurement returns an
answer, then the answer is correct with probability $1$. An
interesting alternative approach for distinguishing between a
collection of quantum states, which is a combination of the
previous two approaches, is to allow for a certain probability of
an inconclusive result, and then maximize the probability of
correct detection \cite{E03p,ZLG99,FJ02}.

We consider a quantum state ensemble consisting of $m$ density
operators $\{\rho_i, 1 \le i \le m\}$ on an $n$-dimensional
complex Hilbert space $\HH$, with prior probabilities $\{p_i>0, 1
\le i \le m\}$. A pure-state ensemble is one in which each density
operator $\rho_i$ is a rank-one projector
$\ket{\phi_i}\bra{\phi_i}$, where the vectors $\ket{\phi_i}$,
though evidently normalized to unit length, are
 not necessarily orthogonal.
Our problem is to design a quantum detector to distinguish
unambiguously between the states $\{\rho_i\}$.

Chefles \cite{C98} showed that a necessary and sufficient
condition for the existence of unambiguous measurements for
distinguishing between a collection of {\em pure} quantum states
is that the states are linearly independent. Necessary and
sufficient conditions on the optimal measurement minimizing the
probability of an inconclusive result for pure states were derived
in \cite{E02}. The optimal measurement when distinguishing between
a broad class of symmetric pure-state sets was also considered in
\cite{E02}.

The problem of unambiguous detection between {\em mixed} state
ensembles has received considerably less attention. Rudolph \etal
\cite{RST03} showed that unambiguous detection between mixed
quantum states is possible as long as one of the density operators
in the ensemble has a non-zero overlap with the intersection of
the kernels of the other density operators. They then consider the
problem of unambiguous detection between two mixed quantum states,
and derive upper and lower bounds on the probability of an
inconclusive result. They also develop a closed form solution for
the optimal measurement in the case in which both states have
kernels of dimension $1$.

In this paper we develop a general framework for unambiguous state
discrimination between a collection of mixed quantum states, which
can be applied to any number of states with arbitrary prior
probabilities.  For our measurement we consider general positive
operator-valued measures \cite{H76,P90}, consisting of $m+1$
measurement operators.  We derive a set of necessary and
sufficient conditions for an optimal measurement that minimizes
the probability of an inconclusive result, by exploiting
principles of duality theory in vector space optimization. We then
show that the previous optimal measurements that were derived for
certain special cases satisfy these optimality conditions.

Next, we consider geometrically uniform (GU) and compound GU state
sets \cite{EF01,EMV02s,EB01}, which are state sets with strong
symmetry properties. We show that the optimal measurement
operators for unambiguous discrimination between such state sets
are also GU and CGU respectively, with generators that can be
computed very efficiently by solving a reduced size semidefinite
program.

The paper is organized as follows. In Section~\ref{sec:problem},
we provide a statement of our problem. In
Section~\ref{sec:conditions} we develop the necessary and
sufficient conditions for optimality using Lagrange duality
theory. Some special cases are considered in Section~\ref{sec:sc}.
In Section~\ref{sec:sym} we consider the problem of distinguishing
between a collection of states with a broad class of symmetry
properties.

\section{Problem Formulation}
\label{sec:problem}

Assume that a quantum channel is prepared in a quantum state drawn
from a collection of mixed states, represented by density
operators $\{ \rho_i,1 \leq i \leq m \}$ on  an $n$-dimensional
complex Hilbert space $\HH$. We assume without loss of generality
that the eigenvectors of $\rho_i,1 \leq i \leq m$, collectively
span\footnote{Otherwise we can transform the problem to a problem
equivalent to the one considered in this paper by reformulating
the problem on the subspace spanned by the eigenvectors of
$\{\rho_i,1 \leq i \leq m\}$. } $\HH$.

To detect the state of the system a measurement is constructed
comprising $m+1$ measurement operators $\{\Pi_i,0 \leq i \leq m\}$
that satisfy
\begin{eqnarray}
\label{eq:psdt}& \Pi_i  \geq  0,\quad 0 \leq i \leq m; &\nonumber
\\
&\sum_{i=0}^m \Pi_i = I.&
\end{eqnarray}
The measurement operators are constructed so that either the state
is correctly detected, or the measurement returns an inconclusive
result. Thus, each of the operators $\Pi_i,1 \leq i \leq m$
correspond to detection of the corresponding states $\rho_i,1 \leq
i \leq m$, and $\Pi_0$ corresponds to an inconclusive result.

Given that the state of the system is $\rho_j$, the probability of
obtaining outcome $i$ is $\tr(\rho_j\Pi_i)$. Therefore, to ensure
that each state is either correctly detected or an inconclusive
result is obtained,  we must have
\begin{equation}
\label{eq:zecond} \tr(\rho_j\Pi_i)=\eta_i\delta_{ij},\quad 1 \leq
i,j \leq m,
\end{equation}
for some $0 \leq \eta_i \leq 1$. Since from (\ref{eq:psdt}),
$\Pi_0=I-\sum_{i=1}^m \Pi_i$, (\ref{eq:zecond}) implies that
$\tr(\rho_i\Pi_0)=1-\eta_i$, so that given that the state of the
system is $\rho_i$, the state is correctly detected with
probability $\eta_i$, and an inconclusive result is returned with
probability $1-\eta_i$.

It was shown in \cite{C98} that for pure-state ensembles
consisting of rank-one density operators
$\rho_i=\ket{\phi_i}\bra{\phi_i}$, (\ref{eq:zecond}) can be
satisfied if and only if the vectors $\ket{\phi_i}$ are linearly
independent. For mixed states, it was shown in \cite{RST03} that
(\ref{eq:zecond}) can be satisfied if and only if one of the
density operators $\rho_i$ has a non-zero overlap with the
intersection of the kernels of the other density operators.
Specifically, denote by $\K_i$ the null space of $\rho_i$ and let
\begin{equation}
\label{eq:Si} \SSS_i=\cap_{j=1,j \neq i}^m\K_j
\end{equation}
 denote the intersection of
$\K_j,1 \leq j \leq m,j \neq i$. Then to satisfy (\ref{eq:zecond})
the eigenvectors of $\Pi_i$ must be contained in $\SSS_i$ and must
not be entirely contained in $\K_i$. This implies that $\K_i$ must
not be entirely contained in $\SSS_i$. Some examples of mixed
states for which unambiguous detection is possible are given in
\cite{RST03}.

 If the state $\rho_i$ is prepared with prior
probability $p_i$, then the total probability of correctly
detecting the state is
\begin{equation}
P_D=\sum_{i=1}^mp_i \tr(\rho_i\Pi_i).
\end{equation}
Our problem therefore is to choose the measurement operators
$\Pi_i,0 \leq i \leq m$ to maximize $P_D$, subject to the
constraints (\ref{eq:psdt}) and
\begin{equation}
\label{eq:zec} \tr(\rho_j\Pi_i)=0,\quad 1 \leq i,j \leq m,i \neq
j.
\end{equation}
To satisfy (\ref{eq:zec}), $\Pi_i$ must lie in $\SSS_i$ defined by
(\ref{eq:Si}), so that
\begin{equation}
\label{eq:pip} \Pi_i=P_i \Pi_iP_i,\quad 1 \leq i \leq m,
\end{equation}
where $P_i$ is the orthogonal projection onto $\SSS_i$. Denoting
by $\Theta_i$ an $n \times r$ matrix whose columns form an
arbitrary orthonormal basis for $\SSS_i$, where $r=\dim(\SSS_i)$,
we can express $P_i$ as $P_i=\Theta_i\Theta_i^*$. From
(\ref{eq:pip}) and (\ref{eq:psdt}) we then have that
\begin{equation}
\Pi_i=\Theta_i\Delta_i\Theta_i^*,\quad 1 \leq i \leq m,
\end{equation}
where $\Delta_i=\Theta_i^* \Pi_i\Theta_i$ is an $r \times r$
matrix satisfying
\begin{eqnarray}
\label{eq:psd}& \Delta_i  \geq  0,\quad 1 \leq i \leq m;
&\nonumber
\\
&\sum_{i=1}^m \Theta_i\Delta_i\Theta_i^* \leq I.&
\end{eqnarray}
Therefore, our problem reduces to maximizing
\begin{equation}
\label{eq:Pd} P_D=\sum_{i=1}^mp_i
\tr(\rho_i\Theta_i\Delta_i\Theta_i^*),
\end{equation}
subject to (\ref{eq:psd}).

To show that the problem of (\ref{eq:Pd}) and (\ref{eq:psd}) does
not depend on the choice of orthonormal basis $\Theta_i$, we note
that any orthonormal basis for $\SSS_i$ can be expressed as the
columns of $\Psi_i$, where $\Psi_i=\Theta_i U_i$ for some $r
\times r$ unitary matrix $U_i$. Substituting $\Psi_i$ instead of
$\Theta_i$ in (\ref{eq:Pd}) and (\ref{eq:psd}), our problem
becomes that of maximizing
\begin{equation}
\label{eq:Pdp} P_D=\sum_{i=1}^mp_i
\tr(\rho_i\Psi_i\Delta_i\Psi_i^*)= \sum_{i=1}^mp_i
\tr(\rho_i\Theta_i\Delta'_i\Theta_i^*),
\end{equation}
where $\Delta_i'=U_i\Delta_iU_i^*$, subject to
\begin{eqnarray}
\label{eq:psdp} & \Delta_i  \geq  0,\quad 1 \leq i \leq m;
&\nonumber
\\
&\sum_{i=1}^m \Psi_i\Delta_i\Psi_i^* =\sum_{i=1}^m
\Theta_i\Delta'_i\Theta_i^*  \leq I.&
\end{eqnarray}
Since $\Delta_i \geq 0$ if and only if $\Delta_i' \geq 0$, the
problem of (\ref{eq:Pdp}) and (\ref{eq:psdp}) is equivalent to
that of (\ref{eq:Pd}) and (\ref{eq:psd}).

Equipped with the standard operations of addition and
multiplication by real numbers, the space $\B$ of all Hermitian $n
\times n$ matrices is an $n^2$-dimensional {\em real} vector
space. As noted in \cite{RST03}, by choosing an appropriate basis
for $\B$, the problem of maximizing $P_D$ subject to
(\ref{eq:psd}) can be put in the form of a standard semidefinite
programming problem, which is a convex optimization problem; for a
detailed treatment of semidefinite programming problems see, \eg
\cite{A91t,A92,NN94,VB96}. By exploiting the many well known
algorithms for solving semidefinite programs \cite{VB96}, \eg
interior point methods\footnote{Interior point methods are
iterative algorithms that terminate once a pre-specified accuracy
has been reached. A worst-case analysis of interior point methods
shows that the effort required to solve a semidefinite program to
a given accuracy grows no faster than a polynomial of the problem
size. In practice, the algorithms behave much better than
predicted by the worst case analysis, and in fact in many cases
the number of iterations is almost constant in the size of the
problem.}  \cite{NN94,A91t}, the optimal measurement can be
computed very efficiently in polynomial time within any desired
accuracy.

Using elements of duality theory in vector space optimization, in
the next section we derive necessary and sufficient conditions on
the measurement operators $\Pi_i=\Theta_i\Delta_i \Theta_i^*$ to
maximize $P_D$ of (\ref{eq:Pd}) subject to (\ref{eq:psd}).

\section{Conditions for optimality}
\label{sec:conditions}
\subsection{Dual Problem Formulation} \label{sec:sdp}

To derive necessary and sufficient conditions for optimality on
the matrices $\Delta_i$  we first derive a dual problem, using
Lagrange duality theory \cite{B99}.

Denote by $\Lambda$  the set of all ordered sets
$\Pi=\{\Pi_i=\Theta_i\Delta_i\Theta_i^*\}_{i=1}^m$ satisfying
(\ref{eq:psd}) and define $J(\Pi)= \sum_{i=1}^m
p_i\tr(\rho_i\Theta_i\Delta_i\Theta_i^*)$. Then our problem is
\begin{equation}
\label{eq:primal} \max_{\Pi \in \Lambda} J(\Pi).
\end{equation}
We refer to this problem as the primal problem, and to any $\Pi
\in \Lambda$ as a primal feasible point. The optimal value of
$J(\Pi)$ is denoted by $\widehat{J}$.

To develop the dual problem associated with (\ref{eq:primal}) we
first compute the Lagrange dual function, which is given by
\begin{eqnarray}
\label{eq:gf} \lefteqn{g(Z)=} \nonumber \\ & \hspace*{-0.3in}= &
\hspace*{-0.1in} \min_{\Delta_i \geq 0}\blc
-\sum_{i=1}^m p_i\tr(\rho_i\Theta_i\Delta_i\Theta_i^*)\right.+ \nonumber \\
&& + \left. \tr\bl Z\bl\sum_{i=0}^m
\Theta_i\Delta_i\Theta_i^*-I\br\br \brc \nonumber \\
& \hspace*{-0.3in}= & \hspace*{-0.1in}\min_{\Delta_i \geq 0}\blc
\sum_{i=1}^m \tr\bl \Delta_i\Theta_i^* \bl Z-p_i\rho_i \br
\Theta_i\br -\tr(Z)\brc,
\end{eqnarray}
where $Z \geq 0$ is the Lagrange dual variable. Since $\Delta_i
\geq 0,1 \leq i \leq m$, we have that $\tr(\Delta_iX) \geq 0$ for
any $X \geq 0$. If $X$ is not positive semidefinite, then we can
always choose $\Delta_i$ such that $\tr(\Delta_i X)$ is unbounded
below. Therefore,
\begin{equation}
g(Z)=\left\{
\begin{array}{ll}
-\tr(Z), & A_i \geq 0, 1 \leq i \leq m, Z \geq 0; \\
-\infty, & \mbox{otherwise},
\end{array}
\right.
\end{equation}
where
\begin{equation}
A_i=\Theta_{i}^*(Z-p_i\rho_i)\Theta_{i},\quad 1\leq i \leq m.
\end{equation}
It follows that the dual problem associated with (\ref{eq:primal})
is
\begin{equation}
\label{eq:dual} \min_{Z} \tr(Z)
\end{equation}
subject to
\begin{eqnarray}
\label{eq:dualc} & \Theta_i^*(Z-p_i\rho_i)\Theta_i\geq 0,
\quad 1 \leq i \leq m; & \nonumber \\
& Z \geq 0. &
\end{eqnarray}
Denoting by $\Gamma$ the set of all Hermitian operators $Z$ such
that $\Theta_{i}^*(Z-p_i\rho_i)\Theta_{i} \geq 0,1 \leq i \leq m$
and $Z \geq 0$, and defining $T(Z)=\tr(Z)$, the dual problem can
be written as
\begin{equation}
\min_{Z \in \Gamma} T(Z).
\label{eq:dualc1}
\end{equation}
We refer to any $Z \in \Gamma$ as a dual feasible point. The
optimal value of $T(Z)$ is denoted by $\widehat{T}$.

\subsection{Optimality Conditions}
\label{sec:opt}
We can immediately verify that both the primal and the dual
problem are strictly feasible. Therefore, their optimal values are
attainable and the duality gap is zero \cite{VB96}, \ie
\begin{equation}
\label{eq:dualitys} \widehat{J}=\widehat{T}.
\end{equation}
In addition, for any
$\Pi=\{\Pi_i=\Theta_i\Delta_i\Theta_i^*\}_{i=1}^m \in \Lambda$ and
$Z \in \Gamma$,
\begin{eqnarray}
\label{eq:wduality} \lefteqn{\hspace*{-0.1in} T(Z)-J(\Pi) =} \nonumber \\
& = &
\tr\bl\sum_{i=1}^m\Theta_{i}\Delta_i\Theta_{i}^*(Z-p_i\rho_i)
+\Pi_0 Z \br  \nonumber \\
& \ge & 0,
\end{eqnarray}
where $\Pi_0=I-\sum_{i=1}^m \Theta_i\Delta_i\Theta_i^* \geq 0$.
Note, that (\ref{eq:wduality}) can be used to develop an upper
bound on the optimal probability of correct detection
$\widehat{J}$. Indeed, since for any $Z \in \Gamma$, $T(Z) \geq
J(\Pi)$, we have that $\widehat{J} \leq T(Z)$ for any dual
feasible $Z$.

Now, let $\hpi_i=\Theta_{i}\widehat{\Delta}_{i}\Theta_{i}^*,1 \leq
i \leq m$ and $\hpi_{0}=I-\sum_{i=1}^m \hpi_i$ denote the optimal
measurement operators that maximize (\ref{eq:Pd}) subject to
(\ref{eq:psd}), and let $\hz$ denote the optimal $Z$ that
minimizes (\ref{eq:dual}) subject to (\ref{eq:dualc}). From
(\ref{eq:dualitys}) and (\ref{eq:wduality}) we conclude that
\begin{equation}
\label{eq:duality}
\tr\bl\sum_{i=1}^m\hpi_i\Theta_{i}^*(\hz-p_i\rho_i)\Theta_{i}+
\hpi_0 \hz \br=0.
\end{equation}
Since $\widehat{\Delta}_i \geq 0$, $\hz \geq 0$ and  $\Theta_{i}^*
(\hz-p_i\rho_i)\Theta_{i} \geq 0,1\leq i \leq m$,
(\ref{eq:duality}) is satisfied if and only if

\begin{eqnarray}
\label{eq:condz1a} & \hz \hpi_0=0 & \\
\label{eq:condz1b} & \Theta_{i}^*(\hz-p_i\rho_i)\Theta_{i}
\widehat{\Delta}_i=0, \quad 1 \leq i \leq m. &
\end{eqnarray}

Once we find the optimal $\hz$ that minimizes
the dual problem (\ref{eq:dual}), the constraints
(\ref{eq:condz1a}) and (\ref{eq:condz1b})
 are necessary and sufficient conditions on the
optimal measurement operators $\hpi_i$. We have already seen that
these conditions are necessary. To show that they are sufficient,
we note that if a set of feasible measurement operators $\hpi_i$
satisfies (\ref{eq:condz1a}) and (\ref{eq:condz1b}), then
$\tr\bl\sum_{i=1}^m\widehat{\Delta}_i\Theta_{i}^*(\hz-p_i\rho_i)\Theta_{i}+
\hpi_0 \hz \br=0$ so that from (\ref{eq:wduality}),
$J(\hpi)=T(\hz)=\widehat{J}$.

We summarize our results in the following theorem:
\begin{theorem}
\label{thm:dual} Let $\{\rho_i,1 \leq i \leq m\}$ denote a set of
density operators with prior probabilities $\{p_i>0,1 \leq i \leq
m\}$, and let $\{\Theta_i,1 \leq i \leq m\}$ denote a set of
matrices such that the columns of $\Theta_{i}$ form an orthonormal
basis for $\SSS_i=\cap_{j=1,j \neq i}^m\K_j$, where $\K_i$ the
null space of $\rho_i$. Let $\Lambda$ denote the set  of all
ordered sets of Hermitian measurement operators
$\Pi=\{\Pi_i\}_{i=0}^m$ that satisfy $\Pi_i \geq 0$, $\sum_{i=0}^m
\Pi_i=I$, and $\tr(\rho_{j}\Pi_{i})=0,1 \leq i \leq m, i\neq j$
and let $\Gamma$ denote the set of Hermitian matrices $Z$ such
that $Z \geq 0$, $ \Theta_{i}^*(Z-p_i\rho_i)\Theta_{i},1 \leq i
\leq m$. Consider the problem $\max_{\Pi \in \Lambda} J(\Pi)$ and
the dual problem $\min_{Z \in \Gamma} T(Z)$, where
$J(\Pi)=\sum_{i=1}^m p_i \tr(\rho_i\Pi_i)$ and $T(Z)=\tr(Z)$. Then
\begin{enumerate}
\item For any $Z \in \Gamma$ and $\Pi \in \Lambda$, $T(Z) \geq J(\Pi)$.
\item There is an optimal $\Pi$, denoted $\hpi$, such that
$\widehat{J}=J(\hpi) \geq J(\Pi)$ for any $\Pi \in \Lambda$;
\item There is an optimal $Z$, denoted $\hz$ and
such that $\widehat{T}=T(\hz) \leq T(Z)$ for any
$Z \in \Gamma$;
\item $\widehat{T}=\widehat{J}$;
\item Necessary and sufficient conditions on the
optimal measurement operators $\hpi_i$ are that there exists a $Z
\in \Gamma$ such that
\begin{eqnarray}
& Z \hpi_0=0 & \\
& \Theta_{i}^*(Z-p_i\rho)\Theta_{i} \widehat{\Delta}_i=0, \quad 1
\leq i \leq m, &
\end{eqnarray}
where $\hpi_i=\Theta_{i}\widehat{\Delta}_i\Theta_{i}^*,1 \leq i
\leq m$, and $\widehat{\Delta}_i \geq 0$.
\item Given $\hz$, necessary and sufficient
conditions on the
optimal measurement operators $\hpi_i$ are
\begin{eqnarray}
& \hz \hpi_0=0 & \\
& \Theta_{i}^*(\hz-p_i\rho_i)\Theta_{i} \widehat{\Delta}_i=0,
\quad 1 \leq i \leq m. &
\end{eqnarray}
\end{enumerate}
\end{theorem}

Although the necessary and sufficient conditions of
Theorem~\ref{thm:dual} are hard to solve, they can be used to
verify a solution and to gain some insight into the optimal
measurement operators. In the next section we show that the
previous optimal measurements that were derived in the literature
for certain special cases satisfy these optimality conditions.

\section{Special cases}
\label{sec:sc}

We now consider two special cases that where addressed in
\cite{RST03}, for which a closed form solution exists. In
Section~\ref{sec:orthog} we consider the case in which the spaces
$\SSS_i$ defined by (\ref{eq:Si}) are orthogonal, and in
Section~\ref{sec:one} we consider the problem of distinguishing
unambiguously between two density operators with $\dim(\SSS_i)=1,1
\leq i \leq 2$.

\subsection{Orthogonal Null Spaces $\SSS_i$}
\label{sec:orthog}

The first case we consider is the case in which the null spaces
$\SSS_i$ are orthogonal, so that
\begin{equation}
P_iP_j=\delta_{ij},\quad 1 \leq i,j, \leq m,
\end{equation}
where $P_i$ is an orthogonal projection onto $\SSS_i$. It was
shown in \cite{RST03} that in this case the optimal measurement
operators are
\begin{equation}
\label{eq:pio} \hpi_{i} = P_i=\Theta_{i}\Theta_{i}^*,\quad 1 \leq
i \leq m.
\end{equation}
In Appendix~\ref{sec:a} we show that the optimal solution of the
dual problem can be expressed as
\begin{equation} \label{eq:zo}
\hz=\sum_{i=1}^{m}p_iP_i\rho_{i}P_i.
\end{equation}
It can easily be shown that $\hz$ and $\hpi_i$ of (\ref{eq:zo})
and (\ref{eq:pio}) satisfy the optimality conditions of Theorem
\ref{thm:dual}.

\subsection{Null Spaces of Dimension $1$}
\label{sec:one}

We now consider the case of distinguishing between two density
operators $\rho_1$ and $\rho_2$, where $\SSS_1$ and $\SSS_2$ both
have dimension equal to $1$. In this case, as we show in Appendix
\ref{sec:b}, the optimal dual solution is
\begin{equation}
\hz=
\begin{cases}
d_{1}P_1, & \text{$d_{2}-d_{1}|f|^2\leq 0$;} \\
d_{2}P_2, & \text{$d_{1}-d_{2}|f|^2\leq 0$;} \\
d_{2}(\Theta_{2}+s\Theta_{2}^{\perp})(\Theta_{2}+s\Theta_{2}^{\perp})^*,
& \text{otherwise,}
\end{cases}
\label{eq:dualf10}
\end{equation}
where $P_i$ is an orthogonal projection onto $\SSS_i$,
$\Theta_2^{\perp}$ is a unit norm vector in the span of $\Theta_1$
and $\Theta_2$ such that $\Theta_{2}^*\Theta_{2}^{\perp}=0$, and
\begin{eqnarray}
\label{eq:def}
& d_i=p_i\Theta_i^*\rho_i\Theta_i, \quad 1 \leq i \leq 2; & \nonumber  \\
&s=\frac{f^*}{e^*}\bl\sqrt{\frac{d_1}{d_2 |f|^2}}-1\br; & \nonumber  \\
& f=\Theta_2^*\Theta_1; & \nonumber \\
& e=(\Theta_2^{\perp})^*\Theta_1. &
\end{eqnarray}
The optimal measurement operators for this case were developed in
\cite{RST03}, and can be written as
\begin{equation}
\{\hpi_i\}_{i=1}^2=
\begin{cases}
\hpi_{1}=P_1,\hpi_{2}=0, &
\text{$d_{2}-d_{1}|f|^2\leq 0$}; \\
\hpi_{1}=0,\hpi_{2}=P_2, &
\text{$d_{1}-d_{2}|f|^2\leq 0$;} \\
\hpi_{1}=\alpha_1P_1,\hpi_{2}=\alpha_2P_2, & \text{otherwise},
\end{cases}
\label{eq:dualf11}
\end{equation}
where
\begin{eqnarray}
& \alpha_1=\frac{1-\sqrt{\frac{d_2|f|^2}{d1}}}{1-|f|^2}; & \nonumber \\
& \alpha_2=\frac{1-\sqrt{\frac{d_1|f|^2}{d2}}}{1-|f|^2}. &
\end{eqnarray}

We now show that $\hz$  and $\hpi$ of (\ref{eq:dualf10}) and
(\ref{eq:dualf11}) satisfy the optimality conditions of
Theorem~\ref{thm:dual}. To this end we note that from
(\ref{eq:dualf11}),
\begin{equation}
\{\widehat{\Delta}_i\}_{i=1}^2=
\begin{cases}
\widehat{\Delta}_{1}=1,\widehat{\Delta}_{2}=0, &
\text{$d_{2}-d_{1}|f|^2\leq 0$}; \\
\widehat{\Delta}_{1}=0,\widehat{\Delta}_{2}=1, &
\text{$d_{1}-d_{2}|f|^2\leq 0$}; \\
\widehat{\Delta}_{1}=\alpha_1,\widehat{\Delta}_{2}=\alpha_2, &
\text{otherwise}.
\end{cases}
\label{eq:delta1}
\end{equation}
From (\ref{eq:dualf10})--(\ref{eq:delta1}) we have that if
$d_{2}-d_{1}|f|^2\leq 0$, then
\begin{eqnarray}
&\Theta_{1}^*(\hz-p_1\rho_1)\Theta_{1}
\widehat{\Delta}_1  =  d_1-\Theta_{1}^*p_1\rho_1\Theta_{1}=0; &\nonumber \\
&\Theta_{2}^*(\hz-p_2\rho_2)\Theta_{2}\widehat{\Delta}_2 = 0; &\nonumber \\
&\hz \hpi_0 =
\hz(I-\hpi_1)=d_1\Theta_1\Theta_{1}^*-d_1\Theta_1\Theta_{1}^*=0. &
\label{eq:satopt1}
\end{eqnarray}
Similarly, if $d_{1}-d_{2}|f|^2\leq 0$, then
\begin{eqnarray}
&\Theta_{1}^*(\hz-p_1\rho_1)\Theta_{1}\widehat{\Delta}_1 = 0; &\nonumber \\
&\Theta_{2}^*(\hz-p_2\rho_2)\Theta_{2}
\widehat{\Delta}_2  =  d_2-\Theta_{2}^*p_2\rho_2\Theta_{2}=0; &\nonumber \\
&\hz \hpi_0 =
\hz(I-\hpi_2)=d_2\Theta_2\Theta_{2}^*-d_2\Theta_2\Theta_{2}^*=0. &
\label{eq:satopt2}
\end{eqnarray}
Finally, if  neither of the conditions $d_{1}-d_{2}|f|^2\leq 0$,
$d_{2}-d_{1}|f|^2\leq 0$ hold, then
\begin{eqnarray}
\lefteqn{\Theta_{1}^*(\hz-p_1\rho_1)\Theta_{1}\widehat{\Delta}_1 =
} \nonumber \\
& = & (d_2(f^*+e^*s)(f^*+e^*s)^*-d_1)\frac{1-\sqrt{\frac{d_2|f|^2}{d_1}}}{1-|f|^2}\nonumber \\
& = & \bl d_2|f|^2\bl\sqrt{\frac{d_1}{d_2|f|^2}}\br^2-d_1 \br
\frac{1-\sqrt{\frac{d_2|f|^2}{d_1}}}
{1-|f|^2} \nonumber \\
&= &0, \label{eq:satopt3}
\end{eqnarray}
\begin{eqnarray}
\Theta_{2}^*(\hz-p_2\rho_2)\Theta_{2}\widehat{\Delta}_2 & = &
(\Theta_2^*\hz
\Theta_2-d_2)\frac{1-\sqrt{\frac{d_1|f|^2}{d_2}}}{1-|f|^2}
\nonumber \\ & = & 0, \label{eq:eqnarray}
\end{eqnarray}
and
\begin{eqnarray}
\hz\hpi_0 &=& \hz-\hz\hpi_{1}-\hz\hpi_{2} \nonumber \\
&=&\hz-\widehat{\Delta}_1\hz
\Theta_1\Theta_1^*-\widehat{\Delta}_2\hz \Theta_2\Theta_2^*.
\label{eq:h}
\end{eqnarray}
To show that $\hz\hpi_0=0$, we note that
\begin{eqnarray}
\hz\Theta_1\Theta_1^*
&=& d_2(|f|^2+s^*ef^*)\Theta_2\Theta_2^* \nonumber \\
&+& d_2(s|f|^2+ss^*ef^*)\Theta_2^{\perp}\Theta_2^* \nonumber \\
&+& d_2(e^*f+s^*|e|^2)\Theta_2\Theta_2^{\perp *} \nonumber \\
&+& d_2(se^*f+ss^*|e|^2)\Theta_2^{\perp}\Theta_2^{\perp *},
\label{eq:h1}
\end{eqnarray}
and
\begin{equation}
\hz\Theta_2\Theta_2^*=d_2\Theta_2\Theta_2^*+d_2s\Theta_2^{\perp}\Theta_2^*.
\label{eq:h2}
\end{equation}
Substituting (\ref{eq:h1}) and (\ref{eq:h2}) into (\ref{eq:h}),
and after some algebraic manipulations, we have that
\begin{equation}
\hz\hpi_0=\hz-\widehat{\Delta}_1\hz
\Theta_1\Theta_1^*-\widehat{\Delta}_2\hz \Theta_2\Theta_2^*=0.
\label{eq:satopt5}
\end{equation}
Combining (\ref{eq:satopt1})--(\ref{eq:satopt5}) we conclude that
the optimal measurement operators of \cite{RST03} satisfy the
optimality conditions of Theorem~\ref{thm:dual}.

\section{Optimal Detection of Symmetric States}
\label{sec:sym}

We now consider the case in which the quantum state ensemble has
symmetry properties referred to as geometric uniformity (GU) and
compound geometric uniformity (CGU). These symmetry properties are
quite general, and include many cases of practical interest.

Under a variety of different optimality criteria the optimal
measurement for distinguishing between GU and CGU state sets was
shown to be GU and CGU respectively \cite{EF01,EMV02s,E02,E03p}.
In particular it was shown in \cite{E02} that the optimal
measurement for unambiguous detection between linearly independent
GU and CGU pure-states is GU and CGU respectively. We now
generalize this result to unambiguous detection of mixed GU and
CGU state sets.

\section{GU State Sets}
\label{sec:gu}

A GU state set is defined as a set of density operators $\{\rho_i,
1 \leq i \leq m\}$ such that $\rho_i=U_i\rho U_i^*$ where $\rho$
is an arbitrary {\em generating operator} and the matrices
$\{U_i,1 \leq i \leq m\}$ are unitary and form an abelian group
$\G$ \cite{F91,EMV02s}. For concreteness, we assume that $U_1=I$.
The group $\G$ is the \emph{generating group} of $\SSS$. For
consistency with the symmetry of $\SSS$, we will assume
equiprobable prior probabilities on $\SSS$.

As we now show, the optimal measurement operators that maximize
the probability of correct detection when distinguishing
unambiguously between the density operators of a GU state set are
also GU with the same generating group. The corresponding
generator can be computed very efficiently in polynomial time.

Suppose that the optimal measurement operators  that maximize
\begin{equation}
J(\{\Pi_i\})=\sum_{i=1}^m \tr(\rho_i \Pi_i)
\end{equation}
subject to (\ref{eq:psd}) and (\ref{eq:zec})
 are $\hpi_i$, and let
$\widehat{J}=J(\{\hpi_i\})=\sum_{i=1}^m\tr(\rho_i \hpi_i)$. Let
$r(j,i)$ be the mapping from $\I \times \I$ to $\I$ with
$\I=\{1,\ldots,m\}$, defined by $r(j,i)=k$ if $U_j^*U_i=U_k$. Then
the measurement operators $\hpi_i^{(j)}=U_j\hpi_{r(j,i)}U_j^*$ and
$\hpi_0^{(j)}=I-\sum_{i=1}^m \hpi_i^{(j)}$ for any $1 \leq j \leq
m$ are also optimal. Indeed, since $\hpi_i \geq 0,1 \leq i \leq m$
and $\sum_{i=1}^m \hpi_i \leq I$, $\hpi^{(j)}_i \geq 0,1 \leq i
\leq m$ and
\begin{equation}
\sum_{i=1}^m \hpi^{(j)}_i=U_j\bl \sum_{i=1}^m \hpi_i \br U_j^*
\leq U_jU_j^*=I.
\end{equation}
Using the fact that $\rho_i=U_i\rho U_i^*$ for some generator
$\rho$,
\begin{eqnarray}
J(\{\hpi^{(j)}_i\})& = & \sum_{i=1}^m\tr(\rho U_i^*
U_j\hpi_{r(j,i)}U_j^*U_i) \nonumber \\
& = & \sum_{k=1}^m\tr(\rho U_k^*\hpi_kU_k) \nonumber \\
& = & \sum_{i=1}^m\tr(\rho_i\hpi_i) \nonumber \\
& = & \widehat{J}.
\end{eqnarray}
Finally, for $l \neq i$,
\begin{eqnarray}
\tr\bl \rho_l\hpi^{(j)}_i \br & = & \tr\bl U_l\rho
U_l^*U_j\hpi_{r(j,i)}U_j^*
\br \nonumber \\
& = & \tr\bl U_s\rho U_s^*\hpi_{r(j,i)} \br \nonumber \\
& = & \tr\bl \rho_s \hpi_k \br\nonumber \\
& = & 0,
\end{eqnarray}
where $U_s=U_j^*U_l$ and $U_k=U_j^*U_i$ and the last equality
follows from the fact that since $l \neq i$, $s \neq k$.

It was shown in \cite{E03p,EMV02s} that if the measurement
operators $\hpi_i^{(j)}$ are optimal for any $j$, then
$\{\overline{\Pi}_i=(1/m)\sum_{j=1}^m \hpi_i^{(j)},1 \leq i \leq
m\}$ and $\overline{\Pi}_0=I-\sum_{i=1}^m \overline{\Pi}_i$ are
also optimal. Furthermore, $\overline{\Pi}_i = U_i
\widehat{\Pi}U_i^*$ where
$\widehat{\Pi}=(1/m)\sum_{k=1}^mU_k^*\hpi_kU_k$.

We therefore conclude that the optimal measurement operators can
always be chosen to be GU with the same generating group $\G$ as
the original state set. Thus, to find the optimal measurement
operators all we need is to find the optimal generator  $\hpi$.
The remaining  operators are obtained by applying the group $\G$
to $\hpi$.

Since the optimal measurement operators satisfy $\Pi_i=U_i \Pi
U_i^*,1 \leq i \leq m$ and $\rho_i=U_i \rho U_i^*$, $\tr(\rho_i
\Pi_i)=\tr (\rho \Pi)$, so that the problem (\ref{eq:Pd}) reduces
to the maximization problem
\begin{equation}
\label{eq:max} \max_{\Pi \in \B} \tr(\rho\Pi),
\end{equation}
where $\B$ is the set of $n \times n$ Hermitian operators, subject
to the constraints
\begin{eqnarray}
\label{eq:condp}
&\Pi  \geq  0; &\nonumber \\
&\sum_{i=1}^m U_i \Pi U_i^* \leq   I; & \nonumber \\
& \tr(\Pi \rho_i)=0, \quad 2 \leq i \leq m. &
\end{eqnarray}
The problem of (\ref{eq:max}) and (\ref{eq:condp}) is a (convex)
semidefinite programming problem, and therefore the optimal $\Pi$
can be computed very efficiently in polynomial time within any
desired accuracy \cite{VB96,A91t,NN94}, for example using the LMI
toolbox on Matlab. Note that the problem of (\ref{eq:max}) and
(\ref{eq:condp}) has $n^2$ real unknowns and $m+1$ constraints, in
contrast with the original maximization problem (\ref{eq:Pd})
subject to (\ref{eq:psd}) and (\ref{eq:zec}) which has $mn^2$ real
unknowns and $m^2+1$ constraints.

\section{CGU State Sets}
\label{sec:cgu}

A CGU state set is defined as a set density operators
$\{\rho_{ik},1 \leq i \leq l,1 \leq k\leq r\}$ such that
$\rho_{ik}=U_i\phi_kU_i^*$ for  some generating density operators
$\{\rho_k,1 \leq k \leq r \}$, where the matrices $\{U_i,1 \leq i
\leq l\}$ are unitary and form an abelian group $\G$
\cite{EB01,EMV02s}. A CGU state set is in general not GU. However,
for every $k$, the operators $\{\rho_{ik},1 \leq i \leq l\}$ are
GU with generating group $\G$.

Using arguments similar to hose of Section~\ref{sec:gu} and
\cite{E03p} we can show that the optimal measurement operators
corresponding to a CGU state set can always be chosen to be GU
with the same generating group $\G$ as the original state set.
Thus, to find the optimal measurement operators all we need is to
find the optimal generators  $\hpi_k$. The remaining  operators
are obtained by applying the group $\G$ to each of the generators
$\hpi_k$.

Since the optimal measurement operators satisfy $\Pi_{ik}=U_i
\Pi_k U_i^*,1 \leq i \leq l,1 \leq k \leq r$ and $\rho_{ik}=U_i
\rho_k U_i^*$, $\tr(\rho_{ik} \Pi_{ik})=\tr (\rho_k \Pi_k)$, so
that the problem (\ref{eq:Pd}) reduces to the maximization problem
\begin{equation}
\label{eq:max2} \max_{\Pi_k \in \B} \sum_{k=1}^r \tr(\rho_k\Pi_k),
\end{equation}
subject to the constraints
\begin{eqnarray}
\label{eq:condp2}
&\Pi_k  \geq  0,\quad 1 \leq k \leq r; &\nonumber \\
&\sum_{i=1}^l \sum_{k=1}^r U_{ik} \Pi_k U_{ik}^* \leq   I; & \nonumber \\
& \tr(\Pi_k \rho_{ik})=0, \quad 1 \leq k \leq r,2 \leq i \leq l. &
\end{eqnarray}
Since this problem is a (convex) semidefinite programming problem,
the optimal generators $\Pi_k$ can be computed very efficiently in
polynomial time within any desired accuracy \cite{VB96,A91t,NN94}.
Note that the problem of (\ref{eq:max2}) and (\ref{eq:condp2}) has
$rn^2$ real unknowns and $lr+1$ constraints, in contrast with the
original maximization which has $lrn^2$ real unknowns and
$(lr)^2+1$ constraints.

\section{Conclusion}

We considered the problem of distinguishing unambiguously between
a collection of {\em mixed} quantum states. Using elements of
duality theory in vector space optimization, we derived a set of
necessary and sufficient conditions on the optimal measurement
operators. We then considered some special cases for which closed
form solutions are known, and showed that they satisfy our
optimality conditions. We also showed that in the case in which
the states to be distinguished have strong symmetry properties,
the optimal measurement operators have the same symmetries, and
can be determined efficiently by solving a semidefinite
programming problem.

An interesting future direction to pursue is to use the optimality
conditions we developed in this paper to derive closed form
solutions for other special cases.

\appendix
\section{Proof of (\ref{eq:zo})}
\label{sec:a}
 To develop the optimal dual solution in the case of orthogonal null spaces, let
$\Theta=\begin{bmatrix}\Theta_1 & \Theta_2 & ... & \Theta_m
\end{bmatrix}$, and define a matrix $\Theta^{\perp}$ such that
$\begin{bmatrix} \Theta & \Theta^{\perp} \end{bmatrix}$ is a
square, unitary matrix, \ie $\begin{bmatrix} \Theta &
\Theta^{\perp}
\end{bmatrix}^*
\begin{bmatrix} \Theta & \Theta^{\perp} \end{bmatrix}=I$. Denoting
$Z={\begin{bmatrix} \Theta & \Theta^{\perp} \end{bmatrix}Y
\begin{bmatrix} \Theta & \Theta^{\perp} \end{bmatrix}^*}$,
the dual problem can be expressed as
\begin{equation}
\min_Y \tr\bl\begin{bmatrix} \Theta & \Theta^{\perp}
\end{bmatrix}Y
\begin{bmatrix} \Theta & \Theta^{\perp} \end{bmatrix}^*\br
\label{eq:p1}
\end{equation}
subject to
\begin{eqnarray}
&\Theta_i^*{\begin{bmatrix} \Theta & \Theta^{\perp} \end{bmatrix}Y
\begin{bmatrix} \Theta & \Theta^{\perp} \end{bmatrix}^*}\Theta_i
\geq \Theta_i^*p_i\rho_i\Theta_i, \quad 1 \leq i \leq m;& \nonumber \\
&Y\geq 0.& \label{eq:p2}
\end{eqnarray}
Using the orthogonality properties of $\Theta_i$ and
$\Theta^\perp$, the problem of (\ref{eq:p1}) and (\ref{eq:p2}) is
equivalent to
\begin{equation}
\min_Y \tr(Y) \label{eq:p3}
\end{equation}
subject to
\begin{eqnarray}
&Y_i
\geq \Theta_i^*p_i\rho_i\Theta_i, \quad 1 \leq i \leq m;& \nonumber \\
&Y\geq 0,& \label{eq:p4}
\end{eqnarray}
where
\begin{equation}
Y=\begin{bmatrix}
Y_1 &  &  &  &\\
 & Y_2 &  &  & \\
 &  & \ddots & &  \\
 &  &  & Y_m & \\
 &  &  & & \text{0}
\end{bmatrix}.
\end{equation}
Since $\tr(Y)=\sum_{i=1}^m\tr(Y_i)$, a solution to (\ref{eq:p3})
subject to (\ref{eq:p4}) is
\begin{equation}
\widehat{Y}=\begin{bmatrix}
\widehat{Y}_1 &  &  &  &\\
 & \widehat{Y}_2 &  &  & \\
 &  & \ddots & &  \\
 &  &  & \widehat{Y}_m & \\
 &  &  & & \text{0}
\end{bmatrix},
\end{equation}
where
\begin{equation}
\widehat{Y}_i=\Theta_i^*p_i\rho_i\Theta_i, \quad 1 \leq i \leq m.
\end{equation}
Then,
\begin{eqnarray}
\hz &=&{\begin{bmatrix} \Theta & \Theta^{\perp} \end{bmatrix}\widehat{Y}
\begin{bmatrix} \Theta & \Theta^{\perp} \end{bmatrix}^*}=
\sum_{i=1}^{m} p_iP_i\rho_i P_i,
\end{eqnarray}
as in (\ref{eq:zo}).

\section{Proof of (\ref{eq:dualf10})} \label{sec:b}

To develop the optimal dual solution $\hz$ for one-dimensional
null spaces, we note that $\hz$ lies in the space spanned by
$\Theta_1$ and $\Theta_2$. Denoting by $\Theta$ a matrix whose
columns represent an orthonormal basis for this space, $\hz$ can
be written as $\hz=\Theta\widehat{Y}\Theta^*$, where the $2 \times
2$ matrix $\widehat{Y}$ is the solution to
\begin{equation}
\min_Y \tr(Y) \label{eq:p11}
\end{equation}
subject to
\begin{eqnarray}
\label{eq:p1a}
&\Phi_1^* Y \Phi_1\geq d_1;& \\
\label{eq:p2a}
&\Phi_2^* Y \Phi_2\geq d_2;& \\
 \label{eq:p3a}
&Y\geq 0.&
\end{eqnarray}
Here $\Phi_i=\Theta^*\Theta_i$ and
$d_i=p_i\Theta_i^*\rho_i\Theta_i$ for $1 \leq i \leq 2$.

To develop a solution to (\ref{eq:p11}) subject to
(\ref{eq:p1a})--(\ref{eq:p3a}), we form the Lagrangian
\begin{equation}
\LL=\tr(Y)-\sum_{i=1}^2\gamma_i (\Phi_i^*Y\Phi_i-d_i)-\tr(XY),
\end{equation}
where from the Karush-Kuhn-Tucker (KKT) conditions \cite{BN01} we
must have that $\gamma_i \geq 0, X \geq 0$, and
\begin{eqnarray}
\label{eq:cs1}
&\gamma_i (\Phi_i^*Y\Phi_i-d_i)=0,\quad i=1,2; \\
\label{eq:cs2} &\tr(XY)=0. &
\end{eqnarray}
Differentiating $\LL$ with respect to $Y$ and equating to zero,
\begin{equation}
I-\sum_{i=1}^2 \gamma_i \Phi_i\Phi_i^*-X=0.
\end{equation}
If $X=0$, then we must have that $I=\sum_{i=1}^2 \gamma_i
\Phi_i\Phi_i^*$, which is possible only if $\Phi_1$ and $\Phi_2$
are orthogonal. Therefore, $X \neq 0$, which implies from
(\ref{eq:cs2}) that (\ref{eq:p3a}) is active. Now, suppose that
only (\ref{eq:p3a}) is active. In this case our problem reduces to
minimizing $\tr(y^*y)$ whose optimal solution is $y=0$, which does
not satisfy (\ref{eq:p1a}) and (\ref{eq:p2a}).

We conclude that at the optimal solution (\ref{eq:p3a}) and at
least one of the constraints (\ref{eq:p1a}) and (\ref{eq:p2a}) are
active. Thus, to determine the optimal solution we need to
determine the solutions under each of the $3$ possibilities: only
(\ref{eq:p1a}) is active, only (\ref{eq:p2a}) is active, both
(\ref{eq:p1a}) and (\ref{eq:p2a}) are active, and then choose the
solution with the smallest objective.

Consider first the case in which (\ref{eq:p1a}) and (\ref{eq:p3a})
are active. In this case, $\widehat{Y}=\hat{y}\hat{y}^*$ for some
vector $\hat{y}$, and without loss of generality we can assume
that
\begin{equation}
\label{eq:yc1} \Phi_1^*\hat{y}=d_1.
\end{equation}
To satisfy (\ref{eq:yc1}), $\hat{y}$ must have the form
\begin{equation}
\label{eq:yf1} \hat{y}=\sqrt{d_1}\Phi_1+\hat{s}\Phi_1^{\perp},
\end{equation}
where $\Phi_1^{\perp}$ is a unit norm vector orthogonal to
$\Phi_1$, so that $\Phi_1^*\Phi_1^{\perp}=0$, and $\hat{s}$ is
chosen to minimize $\tr(\widehat{Y})$. Since,
\begin{equation}
\tr(\widehat{Y})=\hat{y}^*\hat{y}=d_1+|\hat{s}|^2,
\end{equation}
$\hat{s}=0$. Thus, $\widehat{Y}=d_1\Phi_1\Phi_1^*$, and
$\tr(\widehat{Y})=d_1$. This solution is valid only if
(\ref{eq:p2a}) is satisfied, \ie only if
\begin{equation}
\Phi_2^*\widehat{Y}\Phi_2=d_1 |f|^2 \geq d_2.
\end{equation}
Here we used the fact that
\begin{equation}
\Phi_2^*\Phi_1=\Theta_2^*\Theta\Theta^*\Theta_1=\Theta_2^*\Theta_1=f,
\end{equation}
since $\Theta\Theta^*$ is an orthogonal projection onto the space
spanned by $\Theta_1$ and $\Theta_2$.

Next, suppose that (\ref{eq:p2a}) and (\ref{eq:p3a}) are active.
In this case, $\widehat{Y}=\hat{y}\hat{y}^*$ where without loss of
generality we can choose $\hat{y}$ such that
\begin{equation}
\Phi_2^*\hat{y}=d_2,
\end{equation}
and
\begin{equation}
\hat{y}=\sqrt{d_2}\Phi_2+\hat{s}\Phi_2^{\perp},
\end{equation}
where $\Phi_2^{\perp}$ is a unit norm vector orthogonal to
$\Phi_2$, and $\hat{s}$ is chosen to minimize $\tr(\widehat{Y})$.
Since $\tr(\widehat{Y})=d_2+|\hat{s}|^2$, $\hat{s}=0$, and
$\tr(\widehat{Y})=d_2$. This solution is valid only if
(\ref{eq:p1a}) is satisfied, \ie
\begin{equation}
\Phi_1^*Y\Phi_1=d_2 |f|^2 \geq d_1.
\end{equation}

Finally, consider the case in which (\ref{eq:p1a})--(\ref{eq:p3a})
are active. In this case, we can assume without loss of generality
that $\Phi_2^*\hat{y}=\sqrt{d_2}$. Then,
\begin{equation}
\label{eq:haty} \hat{y}=\sqrt{d_2}\Phi_2+\hat{s}\Phi_2^{\perp},
\end{equation}
where $\hat{s}$ is chosen such that
\begin{equation}
\label{eq:ipc} \Phi_{1}^*\widehat{Y}\Phi_{1}=d_1,
\end{equation}
and $\tr(\widehat{Y})=\hat{y}^*\hat{y}$ is minimized. Now, for
$\hat{y}$ given by (\ref{eq:haty}),
\begin{eqnarray}
\lefteqn{\widehat{Y}=d_2\Phi_2\Phi_2^*+
|\hat{s}|^2\Phi_2^{\perp}\Phi_2^{\perp
*}+}\nonumber \\
& &+   \hat{s}\sqrt{d_2}\Phi_2^{\perp}\Phi_2^{*}+
\hat{s}^*\sqrt{d_2}\Phi_2\Phi_2^{\perp *},
\end{eqnarray}
so that
\begin{eqnarray}
\Phi_{1}^*\widehat{Y}\Phi_{1} & = &
d_2|f|^2+|\hat{s}|^2|e|^2+\sqrt{d_2}\hat{s}e^*f+
\sqrt{d_2}\hat{s}^*f^*e \nonumber \\
& = & |\sqrt{d_2}f+\hat{s}^*e|^2, \label{eq:ap1}
\end{eqnarray}
where we defined $\Theta_2^\perp=\Theta \Psi_2^\perp$, and $e$ and
$f$ are given by (\ref{eq:def}). Therefore, to satisfy
(\ref{eq:ipc}), $\hat{s}$ must be of the form
\begin{equation}
\hat{s}=\frac{1}{e^*}\bl e^{j\varphi}\sqrt{d_1}-f^*\sqrt{d_2}\br,
\end{equation}
for some $\varphi$. The problem of (\ref{eq:p11}) then becomes
\begin{equation}
\min_\varphi \frac{1}{|e|^2}\left|
e^{j\varphi}\sqrt{d_1}-f^*\sqrt{d_2}\right|^2,
\end{equation}
which is equivalent to
\begin{equation}
\max_\varphi \Re\left\{{e^{j\varphi}f}\right\}.
\end{equation}
Since
\begin{equation}
\Re\left\{{e^{j\varphi}f}\right\} \leq
\left|e^{j\varphi}f\right|=|f|,
\end{equation}
the optimal choice of $\varphi$ is $e^{j\varphi}=f^*/|f|$, and
\begin{equation}
\hat{s}= \frac{f^*\sqrt{d_2}}{e^*}\bl
\frac{\sqrt{d_1}}{\sqrt{d_2}|f|}-1\br.
\end{equation}
For this choice of $\hat{s}$,
\begin{eqnarray}
\tr(\widehat{Y}) & = & d_2+|\hat{s}|^2 \nonumber \\
& = & d_2\bl 1+\frac{|f|^2}{|e|^2} \bl
\frac{\sqrt{d_1}}{\sqrt{d_2}|f|}-1\br^2 \br \nonumber \\
& \deft & \alpha.
\end{eqnarray}

Clearly, $\alpha \geq d_2$. Therefore, to complete the proof of
(\ref{eq:dualf10}) we need to show that $\alpha \geq d_1$. Now,
\begin{eqnarray}
\lefteqn{|e|^2(\alpha-d_1)  =}\nonumber \\
 & = &
|e|^2(d_2-d_1)+|f|^2 \bl
\frac{\sqrt{d_1}}{|f|}-\sqrt{d_2}\br^2 \nonumber \\
& = & (1-|e|^2)d_1+(|e|^2+|f|^2)d_2-2
\sqrt{d_1}\sqrt{d_2}|f|  \nonumber \\
& = &  (|f|\sqrt{d_1}-\sqrt{d_2})^2 \nonumber \\
& \geq & 0,
\end{eqnarray}
where we used the fact that
\begin{eqnarray}
|e|^2+|f|^2 & = & \Theta_1^*\Theta_2\Theta_2^*\Theta_1+
\Theta_1^*\Theta_2^\perp(\Theta_2^\perp)^*\Theta_1 \nonumber \\
& = & \Theta_1^*\Theta_1=1,
\end{eqnarray}
since $\Theta_2\Theta_2^*+\Theta_2^\perp(\Theta_2^\perp)^*$ is an
orthogonal projection onto the space spanned by $\Theta_1$ and
$\Theta_2$.

\end{document}